\documentclass{eptcs} 

\providecommand{\event}{}
\providecommand{\volume}{??}
\providecommand{\anno}{20??}
\providecommand{\firstpage}{1}

\usepackage{amssymb}
\usepackage{url}
\usepackage{times}
\usepackage{epsfig}
\usepackage{breakurl}

\makeatletter

\def\hml{{\it HML}}
\def\hmli{{\it HML}}
\def\hmlfin{{\it HML}_{\it FIN}}
\def\hmlbd{{\it HML}_{\it FDP}}
\def\hmo{{\cal O}}

\def\nat{\mathbb{N}}
\def\true{{\sf T}}
\def\false{{\sf F}}

\def\hmeq{\sim_{\cal O}}
\def\hmeqone{\sim_{{\cal O}_1}}
\def\hmeqprim{\sim_{{\cal O}'}}

\def\foralli{\forall{i \in I}}

\def\iff{\Leftrightarrow}

\def\implies{\Rightarrow}
\def\nat{\mathbb{N}}
\def\transa{\stackrel{a}{\rightarrow}}

\newtheorem{defi}{Definition}
\newtheorem{theo}{Theorem}
\newtheorem{prop}{Proposition}
\newtheorem{lemm}{Lemma}
\newtheorem{coro}{Corollary}

\newenvironment{theorem}{\begin{theo} \rm }{\end{theo}}
\newenvironment{proposition}{\begin{prop} \rm }{\end{prop}}
\newenvironment{lemma}{\begin{lemm} \rm }{\end{lemm}}
\newenvironment{corollary}{\begin{coro} \rm }{\end{coro}}
\newenvironment{proof}{\begin{trivlist} \item[\hspace{\labelsep}\bf Proof:]}{\hfill$\Box$\end{trivlist}}

\newcommand{\mv}[1]{\mathrel{\stackrel{#1}{\rightarrow}}}
\newcommand{\diam}[1]{\langle#1\rangle}

\title{Modal Logic and the Approximation Induction Principle}
\author{Maciej Gazda \& Wan Fokkink
\institute{Vrije Universiteit\\
Department of Computer Science\\
De Boelelaan 1081a, 1081 HV Amsterdam, Netherlands}
\email{m.w.gazda@student.vu.nl,~wanf@cs.vu.nl}
}
\def\authorrunning{M. Gazda and W. Fokkink}
\def\titlerunning{Modal Characterizations and the Approximation Induction Principle in Process Algebra}

\begin{document}

\maketitle
\thispagestyle{empty}

\begin{abstract}
We prove a compactness theorem in the context of Hennessy-Milner logic.
It is used to derive a sufficient condition on modal characterizations
for the Approximation Induction Principle to be sound modulo the
corresponding process equivalence. We show that this condition
is necessary when the equivalence in question is compositional
with respect to the projection operators.
\end{abstract}

\section{Introduction}

Hennessy-Milner logic \cite{HeMi85} is a modal logic for specifying properties of states in
a labelled transition system (LTS). Rob van Glabbeek \cite{Gla01} uses this logic to
characterize a wide range of process semantics in terms of observations. That is,
a process semantics is captured by means of a sublogic of Hennessy-Milner logic;
two states in an LTS are equivalent if and only if they make true exactly the same
formulas in this sublogic. In particular, Hennessy-Milner logic itself
characterizes bisimulation equivalence.

For several process semantics, mainly in the realm of simulation, van Glabbeek introduces three different modal characterizations (see \cite[Fig.~9]{Gla01}), which differ in their treatment of conjunction. Apart from the richest characterizations, which correspond to the canonical process equivalences, there are also finitary versions (denoted with a superscript $^{*}$), which allow only conjunctions over a finite set. Intermediate equivalences based on formulas with arbitrary conjunctions but finite depth are considered as well (with a superscript $\omega$).  The corresponding equivalences all differ in general LTSs and collapse in the setting of image-finite LTSs. An LTS is image-finite if for each state and each action $a$, there are finitely many outgoing $a$-transitions. Van Glabbeek sketches separate proofs that the modal characterizations capture the same process semantics under consideration. These proofs are always almost identical.

Here we show that given a modal characterization of a process semantics for general LTSs,
restricting to finite sub-conjunctions produces a modal characterization of the same semantics
for image-finite LTSs. The only requirement is that the formulas that are thus obtained
were already present in the original modal characterization.
All semantics in the linear time - branching time spectrum \cite{Gla01} have a modal
characterization that satisfies this requirement, except for completed trace semantics
(in case of an infinite action set).

We obtain a similar compactness result for modal characterizations in which formulas have finite depth.
In this case only infinite conjunctions that have an infinite depth need to
be restricted to their finite sub-conjunctions. Again, the original and the
resulting modal characterization coincide, if the resulting formulas were
already present in the original modal characterization.
The modal characterization of completed trace semantics satisfies this
property. 

Van Glabbeek uses a version of Hennessy-Milner logic that contains negation (so that
disjunction, falsum, and $[a]\,\phi$ need not to be present). However, in that logic
the aforementioned result is not so easy to obtain. Therefore we first prove the result in a
negation-free version of Hennessy-Milner logic. Next we show that the result carries over
to Hennessy-Milner logic with negation. 

Next we study the Approximation Induction Principle (AIP) from process algebra \cite{BaBeKl87},
which states that two processes are equal if they are equal up to any finite depth.
It is well-known that this proof principle is sound modulo bisimulation equivalence for
image-finite processes \cite{Gla87}. Moreover, it is folklore that this soundness
result extends to the other equivalences in the linear time - branching time spectrum
\cite{AcBlVa94}. We obtain a sufficient condition on the modal characterization of a
process equivalence, to guarantee that AIP
is sound with respect to this equivalence.
The result is then linked to the compactness theorem from the first part.
The sufficient condition says that the modal characterization must only
contain formulas of finite depth. We also show that this is basically a
necessary condition: if an equivalence is sound modulo AIP, and
compositional w.r.t.\ the projection operators used in the definition of AIP,
then it can be characterized by a set of finite-depth formulas.

\section{Modal Characterizations for Image-Finite Processes}

\subsection{Hennessy-Milner Logic}

A \emph{labelled transition system} (LTS) consists of a set $S$ of states $s$, a set $A$
of actions $a$, and a set of transitions $s\mv{a}s'$.
An LTS is \emph{image-finite} if for each $s$ and $a$, the LTS contains only finitely
many transitions $s\mv{a}s'$.

Hennessy-Milner logic \cite{HeMi85} is a modal logic for specifying properties of states in an LTS. There exist several versions of Hennessy-Milner logic. The most general language, as presented in \cite{Gla01}, is denoted with $\hmli$. Its syntax can be defined with the following BNF grammar:
\[
\varphi ~~::=~~ \true ~\mid~ \bigwedge_{i\in I}\varphi_i ~\mid~ \diam{a}\varphi ~\mid~ \neg\varphi
\]
The meaning of the formulas is defined inductively as follows:
\[
\begin{array}{l l}
s\models \true &
s\models \diam{a}\varphi ~\Leftrightarrow~ \exists s'\in S\,(s\mv{a}s'\,\wedge\,s'\models\varphi)\\
s\models \bigwedge_{i\in I}\varphi_i ~\Leftrightarrow~ \forall i\in I\,(s\models\varphi_i) ~~~~&
s\models \neg \varphi ~\Leftrightarrow~ s \not \models \varphi
\end{array}
\]
There exists a different syntax (see \cite{Lar90}, \cite{Sti01}) of Hennessy-Milner logic without negation symbol, denoted with $\hmli^{+}$. As we will see later on, its formulas have nice properties which make it easier to perform certain proofs. 
\[
\phi ~~::=~~ \true ~\mid~ \false ~\mid~ \bigwedge_{i\in I}\phi_i ~\mid~ 
\bigvee_{i\in I}\phi_i ~\mid~ \diam{a}\phi ~\mid~ [a]\,\phi
\]
The meaning of the new formulas is defined below:
\[
s\not\models \false \hspace{1cm} 
s\models \bigvee_{i\in I}\phi_i ~\Leftrightarrow~ \exists i\in I\,(s\models\phi_i)
\hspace{1cm} s\models [a]\,\phi ~\Leftrightarrow~ \forall s'\in S\,(s\mv{a}s'\,\Rightarrow\,s'\models\phi)
\]
Observe that we allow quantification over arbitrary sets of indexes $I$. If we restrict to conjunction and disjunction operators over finite sets only, we obtain a language of finite Hennessy-Milner formulas, denoted by $\hmlfin$ or $\hmlfin^+$, respectively.

We define depth of a formula $d:\hmli \longrightarrow \nat \cup \{\infty\}$ inductively as:
\[
d(\true) = 0 \hspace{1cm}
d(\bigwedge_{i\in I}\varphi_i) = \sup\{d(\varphi_i)\mid i\in I\}
\hspace{1cm} d(\diam{a} \varphi) = 1 + d(\varphi)
\hspace{1cm} d(\neg \varphi) = d(\varphi)
\]
$\hmlbd$ and $\hmlbd^+$ denote sets of formulas of finite depth: $\hmlbd^{(+)} = \{ \varphi \in \hmli^{(+)} ~\mid~ d(\varphi) < \infty \}$.

A \emph{context} $C[]$ denotes a formula containing one occurrence of $[]$. The formula
$C[\phi]$ is obtained by replacing this occurrence of $[]$ by the formula $\phi$.
It is well-known, and easy to see, that $\phi\Rightarrow\psi$ yields
$C[\phi]\Rightarrow C[\psi]$ for all contexts $C[]$ over $\hmli^{+}$.

\subsection{Compactness Results}
\label{sec:compactness}

In this section, we show that for image-finite processes, an infinite conjunction or disjunction inside an $\hmli^{+}$ formula
can be captured by its finite sub-conjunctions or -disjunctions, respectively. These results
are somewhat reminiscent of the compactness theorem for first-order logic, which states that
a set of formulas has a model if and only if every finite subset of it has a model.

In \cite{Lar90} there is a result (Lem.~2.8) which implies the proposition below, but only for $\hmlfin^+$ formulas. Moreover, in \cite{Lar90} no proof is provided for Lem.~2.8. Therefore we include a proof of Prop.~\ref{prop:conjunction}, to make the current paper self-contained.

$J\subseteq_{\rm FIN} I$ denotes that $J$ is a finite subset of $I$.

\begin{proposition}
\label{prop:conjunction}
Given an image-finite LTS,
$s\models C[\bigwedge_{i\in I}\phi_i] \in \hmli^{+}$ if and only if
$s\models C[\bigwedge_{i\in J}\phi_i]$ for all $J\subseteq_{\rm FIN} I$.
\end{proposition}

\begin{proof}
$(\Rightarrow)$ For all $J\subseteq_{\rm FIN} I$, $\bigwedge_{i\in I}\phi_i
\Rightarrow \bigwedge_{i\in J}\phi_i$, and so
$C[\bigwedge_{i\in I}\phi_i]\Rightarrow C[\bigwedge_{i\in J}\phi_i]$.

\vspace{4mm}

\noindent
$(\Leftarrow)$ Let $s\models C[\bigwedge_{i\in J}\phi_i]$ for all $J\subseteq_{\rm FIN} I$.
We apply structural induction on $C[]$ to prove that then $s\models C[\bigwedge_{i\in I}\phi_i]$.
\begin{itemize}
\item
$C[]=[]$.

By assumption, $s\models\phi_i$ for all $i\in I$, so $s\models\bigwedge_{i\in I}\phi_i$.

\item
$C[]=C'[]\wedge\bigwedge_{k\in K}\psi_k$.

$s\models C[\bigwedge_{i\in J}\phi_i]$ for all $J\subseteq_{\rm FIN} I$ implies
that $s\models C'[\bigwedge_{i\in J}\phi_i]$ for all $J\subseteq_{\rm FIN} I$, and
$s\models\bigwedge_{k\in K}\psi_k$. By induction, the first fact yields
$s\models C'[\bigwedge_{i\in I}\phi_i]$.
Hence $s\models C[\bigwedge_{i\in I}\phi_i]$.

\item
$C[]=C'[]\vee\bigvee_{k\in K}\psi_k$.

If $s\models\psi_{k_0}$ for some $k_0\in K$, then clearly $s\models C[\bigwedge_{i\in I}\phi_i]$
So suppose $s\models C'[\bigwedge_{i\in J}\phi_i]$ for all $J\subseteq_{\rm FIN} I$.
Then by induction $s\models C'[\bigwedge_{i\in I}\phi_i]$, and so
$s\models C[\bigwedge_{i\in I}\phi_i]$.

\item
$C[]=\diam{a}C'[]$. This is the key case.

By assumption, $s\models\diam{a}C'[\bigwedge_{i\in J}\phi_i]$ for all $J\subseteq_{\rm FIN} I$.
So for each $J\subseteq_{\rm FIN} I$ there is a state $s_J$ such that $s\mv{a}s_J$ and
$s_J\models C'[\bigwedge_{i\in J}\phi_i]$. Since $s$ is image-finite,
$\{s_J\mid J\subseteq_{\rm FIN} I\}$ is finite, say $\{s_{J_1},\ldots,s_{J_m}\}$.
Suppose, towards a contradiction, that $s_{J_k}\not\models C'[\bigwedge_{i\in I}\phi_i]$
for all $k=1,\ldots,m$. Then by induction, for all $k=1,\ldots,m$,
$s_{J_k}\not\models C'[\bigwedge_{i\in K_k}\phi_i]$ for some $K_k\subseteq_{\rm FIN} I$.
This implies that, for all $k=1,\ldots,m$,
$s_{J_k}\not\models C'[\bigwedge_{\ell=1}^m\bigwedge_{i\in K_\ell}\phi_i]$.
This contradicts the fact that $s_{\cup_{\ell=1}^m K_\ell}\in\{s_{J_1},\ldots,s_{J_m}\}$.
We conclude that $s_{J_{k_0}}\models C'[\bigwedge_{i\in I}\phi_i]$
for some $k_0\in\{1,\ldots,m\}$. Hence $s\models \diam{a}C'[\bigwedge_{i\in I}\phi_i]$.

\item
$C[]=[a]\,C'[]$.

Let $s\mv{a}s'$. By assumption, $s'\models C'[\bigwedge_{i\in J}\phi_i]$ for all $J\subseteq_{\rm FIN} I$.
So by induction, $s'\models C'[\bigwedge_{i\in I}\phi_i]$. Hence $s\models[a]\,C'[\bigwedge_{i\in I}\phi_i]$.
\end{itemize}
\end{proof}

\noindent

It is easy to see that Prop.~\ref{prop:conjunction} fails for LTSs that are not image-finite. A counterexample is given at the end of Sect.~\ref{sec:modalchar}. Namely, in that example, the top state at the left does not satisfy $\diam{a}(\bigwedge_{n \in \nat} \diam{a}^{n} \true)$, while it does satisfy $\diam{a}(\bigwedge_{n \in M} \diam{a}^{n} \true)$ for any $M \subseteq_{FIN} \nat$.
\\
\indent There is a counterpart of Prop.~\ref{prop:conjunction}, for disjunction instead of conjunction. To derive this lemma immediately from Prop.~\ref{prop:conjunction}, we introduce an operator that, given a formula in $\hmli^+$, yields a formula equivalent to its negation within $\hmli^+$. Given a $\phi \in \hmli^{+}$, the formula $\overline\phi \in \hmli^{+}$ is defined inductively as follows:

\[
\begin{array}{ccc}
\overline{\sf \true} = \false ~~~~~~~~ &
 \overline{\bigwedge_{i\in I}\phi_i} = \bigvee_{i\in I}\overline{\phi_i} ~~~~~~~~ &
\overline{\diam{a}\phi} = [a]\,\overline{\phi} \vspace{2mm}\\
\overline{\sf \false} = \true ~~~~~~~~ &
\overline{\bigvee_{i\in I}\phi_i} = \bigwedge_{i\in I}\overline{\phi_i} ~~~~~~~~ &
\overline{[a]\,\phi} = \diam{a}\overline{\phi} \\
\end{array}
\]
Clearly, $\neg\phi\,\Leftrightarrow\,\overline{\phi}$. Moreover, $\overline{\overline{\phi}}=\phi$.
The definition is extended to contexts by putting $\overline{[]}=[]$. We write $\overline{C}[]$
for $\overline{C[]}$.
It is easy to see that $\overline{C[\phi]}=\overline{C}[\overline\phi]$.

\begin{proposition}                     
\label{prop:disjunction}
Given an image-finite LTS, $s\models C[\bigvee_{i\in I}\phi_i]$ if and only if
$s\models C[\bigvee_{i\in J}\phi_i]$ for some $J\subseteq_{\rm FIN} I$.
\end{proposition}

\begin{proof}
$
s\models C[\bigvee_{i\in I}\phi_i]~\Leftrightarrow~s\not\models \overline{ C[\bigvee_{i\in I}\phi_i] }
~\Leftrightarrow  s\not\models\overline{C}[\bigwedge_{i\in I}\overline{\phi_i}] 
~\Leftrightarrow~ s\not\models\overline{C}[\bigwedge_{i\in J}\overline{\phi_i}]
\mbox{ for some $J\subseteq_{\rm FIN} I$} \\ \mbox{(by Prop.~\ref{prop:conjunction})}
~\Leftrightarrow~s\not\models \overline{ C[\bigvee_{i\in J}\phi_i] }
\mbox{ for some $J\subseteq_{\rm FIN} I$}
\Leftrightarrow~s\models C[\bigvee_{i\in J}\phi_i]
\mbox{ for some $J\subseteq_{\rm FIN} I$}.
$
\end{proof}

Now we move to Hennessy-Milner logic with negation, $\hmli$. Contexts over this syntax are denoted by $D[]$.
Each formula $\varphi$ over this logic can be translated to an equivalent formula
${\it P}(\varphi)\in\hmli^+$ in a straightforward fashion:
\[
\begin{array}{lcllcl}
{\it P}(\true) &=& \true ~~~~&
~~~~{\it P}(\bigwedge_{i\in I}\varphi_i) &=& \bigwedge_{i\in I}{\it P}(\varphi_i)\\
{\it P}(\diam{a}\varphi) &=& \diam{a}{\it P}(\varphi)~~~~&
~~~~{\it P}(\neg\varphi) &=& \overline{{\it P}(\varphi)}
\end{array}
\]
Clearly, $\varphi\Leftrightarrow P(\varphi)$.
The definition is extended to contexts by putting ${\it P}([])=[]$. We write ${\it P}(D)[]$ for
${\it P}(D[])$.

For Hennessy-Milner logic with negation, we inductively define \emph{positive} and \emph{negative}
contexts as follows.
\begin{itemize}
\item
$[]$ is a positive context.
\item
If $D[]$ is a positive (resp.\ negative) context, then $D[]\wedge\bigwedge_{i\in I}\varphi_i$
and $\diam{a}D[]$ are positive (resp.\ negative) contexts.
\item
If $D[]$ is a positive (resp.\ negative) context, then $\neg D[]$ is a negative (resp.\ positive)
context.
\end{itemize}

\begin{lemma}
\label{lem:contexts}
${\it P}(D[\varphi])={\it P}(D)[{\it P}(\varphi)]$ if $D[]$ is a positive context, and
${\it P}(D[\varphi])={\it P}(D)[\overline{{\it P}(\varphi)}]$ if $D[]$ is a negative context.
\end{lemma}

\begin{proof}
We prove both statements simultaneously,
by structural induction on $D[]$. The cases where $D[]$ is of the form $[]$, $D'[]\wedge\bigwedge_{i\in I}\varphi_i$
or $\diam{a}D'[]$ are straightforward and left to the reader. We focus on the key case $D[]=\neg D'[]$.

First let $D[]$ be positive, so $D'[]$ is negative. Then
${\it P}(\neg D'[\varphi])=\overline{{\it P}(D'[\varphi])}=\overline{{\it P}(D')[\overline{{\it P}(\varphi)}]}\mbox{ (by induction)}\\
=\overline{{\it P}(D')}[\overline{\overline{{\it P}(\varphi)}}]={\it P}(\neg D')[{\it P}(\varphi)]$.

Next let $D[]$ be negative, so $D'[]$ is positive. Then
${\it P}(\neg D'[\varphi])=\overline{{\it P}(D'[\varphi])}=\overline{{\it P}(D')[{\it P}(\varphi)]}\mbox{ (by induction)}\\
=\overline{{\it P}(D')}[\overline{{\it P}(\varphi)}]={\it P}(\neg D')[\overline{{\it P}(\varphi)}]$.
\end{proof}

\noindent
Now we can prove a counterpart of Prop.~\ref{prop:conjunction} and \ref{prop:disjunction}
for $\hmli$.

\begin{proposition}
\label{prop:negation}
Given an image-finite LTS.
\begin{enumerate}
\item
If $D[]$ is a positive context, then
$s\models D[\bigwedge_{i\in I}\varphi_i]$ if and only if
$s\models D[\bigwedge_{i\in J}\varphi_i]$ for all $J\subseteq_{\rm FIN} I$.
\item
If $D[]$ is a negative context, then
$s\models D[\bigwedge_{i\in I}\varphi_i]$ if and only if
$s\models D[\bigwedge_{i\in J}\varphi_i]$ for some $J\subseteq_{\rm FIN} I$.
\end{enumerate}
\end{proposition}

\begin{proof}
If $D[]$ is a positive context, then\\
$s\models D[\bigwedge_{i\in I}\varphi_i]~\Leftrightarrow~s\models {\it P}(D[\bigwedge_{i\in I}\varphi_i])~\Leftrightarrow
s\models {\it P}(D)[\bigwedge_{i\in I}{\it P}(\varphi_i)]\mbox{ (by Lem.~\ref{lem:contexts})}~\Leftrightarrow\\
s\models {\it P}(D)[\bigwedge_{i\in J}{\it P}(\varphi_i)]\mbox{ for all $J\subseteq_{\rm FIN} I$ (by Prop.~\ref{prop:conjunction})}
\Leftrightarrow~s\models {\it P}(D[\bigwedge_{i\in J}\varphi_i])\mbox{ for all $J\subseteq_{\rm FIN} I$ (by Lem.~\ref{lem:contexts})}\\
\Leftrightarrow~s\models D[\bigwedge_{i\in J}\varphi_i]\mbox{ for all $J\subseteq_{\rm FIN} I$}$.

\vspace{2mm}

\noindent
If $D[]$ is a negative context, then\\
$s\models D[\bigwedge_{i\in I}\varphi_i]~\Leftrightarrow~s\models {\it P}(D[\bigwedge_{i\in I}\varphi_i])~\Leftrightarrow
s\models {\it P}(D)[\bigvee_{i\in I}\overline{{\it P}(\varphi_i)}]\mbox{ (by Lem.~\ref{lem:contexts})}~\Leftrightarrow\\
s\models {\it P}(D)[\bigvee_{i\in J}\overline{{\it P}(\varphi_i)]}\mbox{ for some $J\subseteq_{\rm FIN} I$ (by Prop.~\ref{prop:disjunction})}
\Leftrightarrow~s\models {\it P}(D[\bigwedge_{i\in J}\varphi_i])\mbox{ for some $J\subseteq_{\rm FIN} I$ (by Lem.~\ref{lem:contexts})}\\
\Leftrightarrow~s\models D[\bigwedge_{i\in J}\varphi_i]\mbox{ for some $J\subseteq_{\rm FIN} I$}$.
\end{proof}

\subsection{Modal Characterizations}
\label{sec:modalchar}

A process semantics on LTSs can be captured by means of a sublogic of $\hmli$; see \cite{Gla01} for a wide range of such modal characterizations.
Given such a sublogic ${\cal O}$, two states in an LTS are equivalent if and only if they make true
exactly the same formulas in ${\cal O}$. We denote this equivalence relation on states by $\sim_{\cal O}$.

We will prove that given such a modal characterization of a process semantics for general LTSs,
restricting infinite conjunctions to their finite sub-conjunctions produces a modal characterization
of the same semantics, on image-finite LTSs. The only requirement is that these finite
sub-conjunctions are already present in the original modal characterization for general LTSs.

We obtain a similar compactness result for modal characterizations of which the
formulas may contain infinite conjunctions, but are all of finite depth.
In this case only infinite conjunctions that have an infinite depth need to
be restricted to their finite sub-conjunctions. Again, the original and the
resulting modal characterization coincide, if the resulting formulas were
already present in the original modal characterization.

The modal characterizations in \cite{Gla01} all satisfy this requirement,
except for the one of completed trace semantics, in case of an infinite action set.
Namely, the modal characterization of completed trace semantics, for general processes as well as for
image-finite ones, is:
\[
\varphi~::=~\true~\mid~\bigwedge_{a\in A}\neg\diam{a}\true~\mid~\diam{a}\varphi
\]
where $A$ denotes the set of all actions.

Given a modal characterization ${\cal O}$, we denote the sublogic of formulas in ${\cal O}$
that do not contain infinite conjunctions by ${\cal O}_{\rm FIN}$ and the sublogic of formulas with finite depth with ${\cal O}_{FDP}$. Clearly ${\cal O}_{\rm FIN} \subseteq {\cal O}_{FDP}$. Using the results from Sect.\
\ref{sec:compactness}, we can now prove the main theorem of this section.

\begin{theorem}
\label{thm:hml}
Given an image-finite LTS, and ${\cal O}\subseteq\hmli$.
\begin{enumerate}
\item If for each $D[\bigwedge_{i\in I}\varphi_i]\in{\cal O}$ with $I$ infinite and $d(\bigwedge_{i\in I}\varphi_i) = \infty$,
$D[\bigwedge_{i\in J}\varphi_i]\in{\cal O}$ for all $J\subseteq_{\rm FIN} I$, 
then $\sim_{\cal O}$ and $\sim_{{\cal O}_{\rm FDP}}$ coincide.

\item If for each $D[\bigwedge_{i\in I}\varphi_i]\in{\cal O}$ with $I$ infinite, $D[\bigwedge_{i\in J}\varphi_i]\in{\cal O}$ for all $J\subseteq_{\rm FIN} I$,
then $\sim_{\cal O}$ and $\sim_{{\cal O}_{\rm FIN}}$ coincide.
\end{enumerate}
\end{theorem}

\begin{proof}
We will prove the theorem for the subset of finite formulas ${\cal O}_{\rm FIN}$, and make remarks between square brackets about the version with ${\cal O}_{\rm FDP}$ whenever it is necessary. Since ${\cal O}_{\rm FIN}\subseteq{\cal O}_{\rm FDP}\subseteq {\cal O}$, clearly
$\sim_{\cal O}\,\subseteq\,\sim_{{\cal O}_{\rm FDP}}\,\subseteq\,\sim_{{\cal O}_{\rm FIN}}$. We need to show that
${\cal O}_{\rm FIN}$ [resp.\ ${\cal O}_{\rm FDP}$] can distinguish all states that ${\cal O}$ can.

Given states $s,s'$ and a formula $\varphi\in{\cal O}$ with $s\models\varphi$
and $s'\not\models\varphi$. We will construct a formula in ${\cal O}_{\rm FIN}$ [resp.\ ${\cal O}_{\rm FDP}$]
that distinguishes $s$ and $s'$. We apply ordinal induction on the length $\lambda(\varphi)$
of the longest chain of nested infinite conjunctions [of infinite depth] in $\varphi$. That is,
\[
\begin{array}{lcl}
\lambda(\true) &=& 0 \\
\lambda(\diam{a}\varphi) &=& \lambda(\varphi) \\
\lambda(\bigwedge_{i\in I}\varphi_i) &=& \left\{
\begin{array}{ll}
1+\sup\{\lambda(\varphi_i)\mid i\in I\} & \mbox{if $I$ is infinite [and $d(\bigwedge_{i\in I}\varphi_i)=\infty$]} \\
\sup\{\lambda(\varphi_i)\mid i\in I\} & \mbox{otherwise}
\end{array}
\right.\\
\lambda(\neg \varphi) &=& \lambda(\varphi)
\end{array}
\]
The base case is trivial, because if $\lambda(\varphi)=0$, then $\varphi\in{\cal O}_{\rm FIN}$ [resp.\ $\varphi\in{\cal O}_{\rm FDP}$].
Now consider the inductive case, where $\lambda(\varphi)>0$.
Let $\varphi=D[\bigwedge_{i\in I}\varphi_i]$ with $I$ [and $d(\bigwedge_{i\in I}\varphi_i)$] infinite, where this occurrence of an infinite
conjunction [and depth] in $\varphi$ is outermost, in the sense that it does not occur within any infinite
conjunction [of infinite depth]. We distinguish two cases.
\begin{itemize}
\item
$D[]$ is a positive context. By Prop.~\ref{prop:negation}.1, $s'\not\models\varphi$
implies that $s'\not\models D[\bigwedge_{i\in {J_0}}\varphi_i]$ for some
$J_0\subseteq_{\rm FIN} I$, while $s\models\varphi$ implies that
$s\models D[\bigwedge_{i\in {J_0}}\varphi_i]$.
\item
$D[]$ is a negative context. By Prop.~\ref{prop:negation}.2, $s\models\varphi$
implies that $s\models D[\bigwedge_{i\in {J_0}}\varphi_i]$ for some
$J_0\subseteq_{\rm FIN} I$, while $s'\not\models\varphi$ implies that
$s'\not\models D[\bigwedge_{i\in {J_0}}\varphi_i]$.
\end{itemize}
In both cases, by assumption, $D[\bigwedge_{i\in {J_0}}\varphi_i]\in{\cal O}$.

Clearly, there are only finitely many outermost occurrences of infinite conjunctions
[of infinite depth] in $\varphi$. Using the construction above, these can all be
replaced by finite conjunctions, to obtain a formula $\psi\in{\cal O}$ that
distinguishes $s$ and $s'$. Since $\lambda(\psi)<\lambda(\varphi)$, by ordinal induction,
we can construct a formula in ${\cal O}_{\rm FIN}$ [resp. ${\cal O}_{\rm FDP}$] that distinguishes $s$ and $s'$.
\end{proof}

\noindent
It is easy to see that the requirement in Thm.~\ref{thm:hml} that
$D[\bigwedge_{i\in J}\varphi_i]\in{\cal O}$ for all $J\subseteq_{\rm FIN} I$
cannot be omitted. For instance, let ${\cal O}$ consist of a single formula
with an infinite conjunction, $\bigwedge_{n\in{\mathbb N}}\diam{a}^n{\sf \true}$
(with $\diam{a}^0\varphi=\varphi$ and $\diam{a}^{n+1}\varphi=\diam{a}(\diam{a}^n\varphi)$).
Then ${\cal O}_{\rm FIN}=\emptyset$, so $\sim_{{\cal O}_{\rm FIN}}$ is the universal
relation. On the other hand, ${\cal O}$ distinguishes an $a$-cycle from a deadlock state.

The following example, taken from \cite{Gla87}, shows that Thm.~\ref{thm:hml}
fails for LTSs that are not image-finite. Consider an LTS that consists of finite
$a$-traces of arbitrary length, and an LTS that on top of this exhibits an
infinite $a$-trace.

\vspace{3mm}

\begin{center}
\includegraphics[width=80mm]{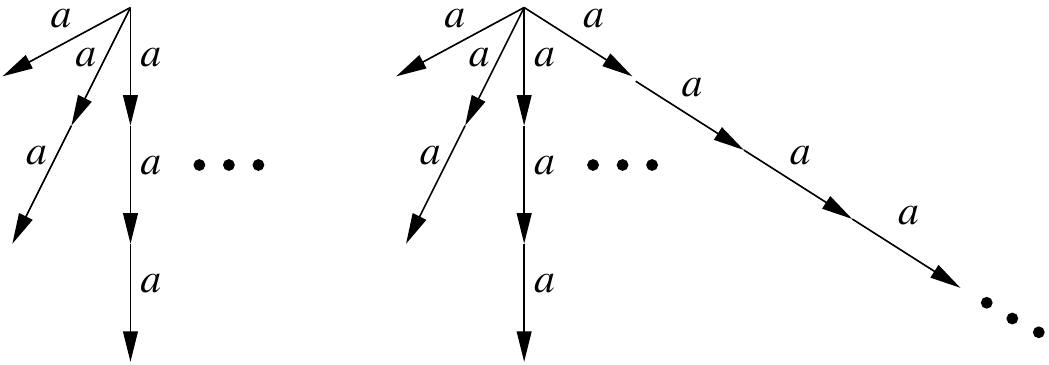}
\end{center}

\vspace{3mm}

\noindent
Let ${\cal O}=\{\diam{a}(\bigwedge_{n\in N}\diam{a}^n{\sf \true})\mid N\subseteq{\mathbb N}\}$.
Then ${\cal O}_{\rm FIN}=\{\diam{a}(\bigwedge_{n\in N}\diam{a}^n \true)\mid N\subseteq_{\rm FIN}{\mathbb N}\}$.
Clearly, ${\cal O}$ distinguishes the top states of the two LTSs above, by means of
any formula $\diam{a}({\bigwedge_{n\in N}\diam{a}^n\true})$ with $N$ infinite.
Namely, such a formula holds for the top state at the right, but not for the top
state at the left. However, ${\cal O}_{\rm FIN}$ does not distinguish these states;
all formulas in ${\cal O}_{\rm FIN}$ hold for both states.

Goldblatt \cite{Gol95} and Hollenberg \cite{Hol95} (see also \cite{BlRiVe01}) investigated models that are more general than image-finite LTSs, but that do have the Hennessy-Milner property. That is, models where the modal equivalence $\sim_{\hml}$ coincides with bisimulation equivalence. This led to the notion of modally saturated processes; an LTS is M-saturated if for all states $s$ and all ${\cal O}\subseteq\hml$, whenever every finite subset of ${\cal O}$ is satisfied in some $a$-successor of $s$, then there exists an $a$-successor of $s$ in which ${\cal O}$ is satisfied. It is not difficult to prove, with ordinal induction on the structure of formulas, that Thm.~\ref{thm:hml} holds for M-saturated models as well.

\section{Approximation Induction Principle}

For each natural number $n$ we define a \textit{projection operator} $\pi_{n}$ which mimicks the behaviour of its argument up to $n$ steps and then terminates. The behaviour of an application of the projection operator to a process (or state) is given by the following rule scheme:
\begin{center}
\Large
$\frac{~~~x \stackrel{a}{\rightarrow}x'}{\pi_{n+1}(x)\stackrel{a}{\rightarrow}\pi_{n}(x')}$
\end{center}
The \textit{Approximation Induction Principle (AIP)} states that if two processes are equal up to any finite depth, then the processes themselves are equal.
\begin{center}
(AIP) If $\pi_{n}(x) = \pi_{n}(y)$ for all $n \in N$, then $x = y$. 
\end{center}

\subsection{Sufficient Criterion for Soundness of AIP}

In \cite{AcBlVa94} it is stated that AIP is sound for all 11 "strong" equivalences from \cite{Gla01}, but no argument is provided.
Soundness of AIP has been proved several times for bisimulation equivalence (e.g. \cite{Gla87}) in the setting of finitely branching or image-finite processes. The standard technique is to prove that a relation identifying two processes if and only if all of their projections are bisimilar is a bisimulation (provided that one of the processes is image-finite). A different proof has been presented in \cite{BaWe90}. Given two processes $p$ and $q$ the authors consider, for all $n \in \nat$, the bisimulations between $\pi_n(p)$ and $\pi_n(q)$. Bisimulations for $n$-th projection are linked with those bisimulations for $(n{+}1)$-th projection in which they are included. This way an infinite, finitely branching tree is constructed. The bisimulation between $p$ and $q$ is a sum of bisimulations lying on an infinite path in the tree.

We present a general proof of soundness of AIP in a different way for a range of equivalences, using properties of modal languages that define an equivalence. Namely, AIP is sound for all process equivalences that can be defined using modal characterizations within $\hmlbd$. The crucial part of the proof is the following lemma which states that if a finite-depth formula is satisfied by a process, then it is satisfied by almost all of its projections. 

\begin{lemma}
\label{lem:boundedform}
Given any LTS, for all states $s$ and $\varphi \in \hmlbd$:
\begin{center}
 $s \models \varphi$ $\Leftrightarrow$ $\forall n \geq d(\varphi) ~ \pi_{n}(s) \models \varphi$
\end{center}
\end{lemma}
 
\begin{proof}
Let $s$ be an arbitrary state. We will proceed with induction on the complexity of a formula, defined by:\\ 
\[
|\true| = 1
\hspace{1cm} |\diam{a} \varphi| = 1 + |\varphi|
\hspace{1cm} |\bigwedge_{i\in I}\varphi_i| = 1 + \sup\{|\varphi_i|\mid i\in I\}
\hspace{1cm} |\neg \varphi| = 1 + |\varphi|
\]
"$\Rightarrow$": The base is trivial ($\varphi = \true$). Let $\varphi$ be a formula such that $s \models \varphi$, and suppose that for all $s'$ and for all $\psi$ with $|\psi| < |\varphi|$, $s' \models \psi$ implies that $\psi$ is satisfied by all projections $\pi_n(s')$ for $n \geq d(\psi)$. There are three possible cases:
\begin{itemize}
	\item $\varphi = \diam{a} \psi$\\
	Then $\exists q$: $s \stackrel{a}{\rightarrow} q \wedge q \models \psi$ with $q \models \psi$.
	From the induction hypothesis we obtain: $\forall n \geq d(\psi) ~~ \pi_{n}(q) \models \psi$. Since
	$\pi_{n}(s) \stackrel{a}{\rightarrow} \pi_{n-1}(q)$ for $n \geq 1$, we have: $\forall n \geq d(\psi)+1$ $\pi_{n}(s) \models \diam{a} \psi$, so $\forall n \geq d(\diam{a} \psi)$ $\pi_{n}(s) \models \diam{a} \psi$ 
	
	\item $\varphi = \bigwedge_{i \in I} \psi _{i}$\\
	Then $~\forall {i \in I} s \models \psi _{i}$.
	By induction, this implies: $\forall i \in I$ $\forall n \geq d(\psi_{i})$ $\pi_{n}(s) \models \psi_{i}$.
	Therefore	$\forall n \geq max_{i \in I} \{ d(\psi_i)\}$, $\forall i \in I$ $\pi_{n}(s) \models \psi_{i}$.
	By definition $d(\bigwedge_{i \in I} \psi_i) = max_{i \in I} \{ d(\psi_i)\}$, so	$\forall n \geq d(\bigwedge_{i \in I} \psi_i)$ $\pi_{n}(s) \models \bigwedge_{i \in I} \psi_{i}$.
	
	\item $\varphi = \neg \psi$\\
	We have to consider all the subcases, depending on $\psi$:\\
	- $\psi = \true$: this is impossible (it would mean that $s \not \models \true$ which is never true).
	\\
	- $\psi = \diam{a} \psi'$: Then $\forall s': s\stackrel{a}{\rightarrow}s'$ we have $s' \models \neg \psi'$. By induction
	$\forall s': s \transa s'$ we have $\forall n \geq d(\neg \psi')$ $\pi_n(s') \models \neg \psi '$. Therefore
	$\forall n \geq d(\neg \psi')+1 $ $\pi_{n}(s) \models \neg \diam{a} \psi '$,
	and thus
	$\forall n \geq d(\psi) $ $\pi_{n}(s) \models \neg \diam{a} \psi '$.\\
	- $\psi = \bigwedge_{i \in I} \psi_{i}$: Then
	$\exists i_{0} \in I: s \models \neg \psi_{i_{0}}$.
  By induction,
  $\forall n \geq d(\neg \psi_{i_{0}})$: $\pi_{n}(s) \models \neg \psi_{i_{0}}$,
  and so $\forall n \geq d(\varphi) \pi_{n}(s) \models \neg \bigwedge_{i \in I} \psi_{i}$, which is the desired statement.	
  \\
  - $\psi = \neg \psi '$: This is immediate (in this case $\varphi$ is equivalent to $\psi '$).
\end{itemize}
"$\Leftarrow$": The other direction follows immediately from what we have just proven. Take an arbitrary formula $\varphi \in \hmo$ and a state $s$ such that
$\forall n \geq d(\varphi) ~ \pi_n(s) \models \varphi$. Suppose towards a contradiction that $s \not \models \varphi$. Then $s \models \neg \varphi$, and it was already proven that this implies $\forall n \geq d(\neg \varphi) ~ \pi_n(s) \models \neg \varphi$. This contradicts our assumptions. Therefore $s$ must satisfy $\varphi$.
\end{proof}

\begin{theorem}
\label{thm:aip}
If $\hmo \subseteq \hmlbd$, then AIP is sound for $\hmeq$.
\end{theorem}

\begin{proof}
We need to show that 
$\forall n \in N\,(\pi_{n}(s) \hmeq \pi_{n}(q))$ $\Rightarrow$ $s \hmeq q$. 
Suppose that $\forall n \in N\,(\pi_{n}(s) \hmeq \pi_{n}(q))$.
We have to prove that $\hmo(s) = \hmo(q)$. In fact it suffices to prove that $\hmo(s) \subseteq \hmo(q)$, the proof of the other inclusion is symmetric. Take any $\varphi \in \hmo(s)$. According to the Lemma \ref{lem:boundedform}, $\forall n \geq d(\varphi)$ $\varphi \in \hmo(\pi_{n}(s)) = \hmo(\pi_{n}(q))$. Using the same lemma again we obtain $\varphi \in {\cal O}(q)$.
\end{proof}

In view of the results from the previous section, we obtain the following sufficient condition for the soundness of AIP in the setting of image-finite LTSs.

\begin{corollary}
\label{cor:aip1}
Let ${\cal O}\subseteq\hmli$. Suppose that for each $D[\bigwedge_{i\in I}\varphi_i]\in{\cal O}$ with $I$ infinite and $d(\bigwedge_{i\in I}\varphi_i) = \infty$,~ $D[\bigwedge_{i\in J}\varphi_i]\in{\cal O}$ for all $J\subseteq_{\rm FIN} I$. Then AIP is sound for $\hmeq$ in the setting of image-finite processes.
\end{corollary}

\begin{proof}
If $\hmo$ meets the above requirements, then according to Thm.~\ref{thm:hml}.2 $~\hmeq = \hmeqprim$, where $\hmo' \in \hmlbd$. By Thm.~\ref{thm:aip} AIP is sound for $\hmeq$.
\end{proof}

\begin{corollary}
\label{cor:aip2}
AIP is sound with respect to all the basic process equivalences on image-finite processes, namely  trace, completed trace, failures, readiness, failure trace, ready trace, ready simulation, $n$-nested simulation ($n \geq 1$), bisimulation.
\end{corollary}

\begin{proof}
As pointed in \cite{Gla01}, all the above equivalences with the exception of completed trace can be defined with a sublogic of Hennessy-Milner logic consisting of finite formulas. Moreover, all formulas in the modal language corresponding to completed trace equivalence are finite-depth.
\end{proof}

Soundness of AIP does not necessarily imply that the equivalence in question is definable with a sublogic of $\hmlbd$. Observe first that having a fixed set of actions $A$, for any formula $\varphi \in \hmli$  we can express an ACTL formula $E \varphi$ ("there exists an execution path to a state in which $\varphi$ holds") with a single formula from $\hmli$. Indeed, for any $\varphi \in \hmli$ the formula $\bigvee_{\sigma \in A^{*}} \sigma \varphi$ is equivalent to $E \varphi$. Now consider an equivalence relating two processes according to whether action $a$ can be executed in at least one execution path (that is, if $E (\diam{a} \true)$ is satisfied). It is easy to observe that AIP is sound for this equivalence, but it cannot be defined with a sublogic of $\hmlbd$.

\subsection{Necessary Criterion for Soundness of AIP}

In this section we consider only those equivalences which are compositional w.r.t.\ projection operators (this includes all the equivalences mentioned in Corollary ~\ref{cor:aip2}) . We will prove that in this class, definability of an equivalence with finite-depth formulas is also a  necessary condition for the soundness of AIP. 

First we define for each $\varphi \in \hmli$ a corresponding formula $cut_n(\varphi) \in \hmlbd$ in which every subformula of the form $\diam{a}\psi$ appearing at depth $n$ is replaced with $\false$.
The functions $cut_{n}: \hmli \rightarrow \hmlbd $ for $n \in \nat$ are defined inductively as follows:
\[
\begin{array}{l l l}
cut_n(\true) = \true &
cut_0(\diam{a} \varphi) = \false &
cut_n(\neg \varphi) = \neg cut_n(\varphi) \\
cut_n(\bigwedge_{i \in I} \varphi_i) = \bigwedge_{i \in I} cut_n(\varphi_i)~~~~ &
cut_{n+1}(\diam{a}\varphi) = \diam{a}cut_n(\varphi)~~~~
\end{array}
\]
We now prove a key property for $cut$ functions.

\begin{lemma}
\label{lem:cut}
Given any LTS. For all states $s$, $\varphi \in \hmli$ and $n \in \nat$:
\begin{center}
(CT)~~ $\pi_n(s) \models \varphi \iff  s \models cut_n(\varphi)$
\end{center}
\end{lemma}

\begin{proof}
We prove CT by induction on the structure of $\varphi$.
\begin{itemize}
\item $\varphi = \true$:
\\
$\pi_n(s) \models \true$  and $s\models cut_n(\true)=\true$.

\item $\varphi = \diam{a} \psi$: Now we distinguish cases where $n=0$ and $n > 0$.

\vspace{2mm}

Clearly $\pi_0(s) \not\models \diam{a} \psi$ and $s\not\models cut_0(\diam{a} \psi)=\false$.

\vspace{2mm}
If $ n > 0$, then
$\pi_{n}(s) \models \diam{a} \psi$ ~
$\iff \exists s': s \transa s' \wedge \pi_{n-1}(s') \models \psi$ (transition rules for $\pi_{n-1}$)~
$\iff \exists s': s \transa s'$ \\ $\wedge~ s' \models cut_{n-1}(\psi)$ (structural induction) ~
$\iff s \models \diam{a}cut_{n-1}(\psi)$~
$\iff s \models cut_{n}(\diam{a} \psi)$ (definition of $cut$)

\item $\varphi = \bigwedge_{i \in I} \psi_i$:\\
$\pi_n(s) \models \bigwedge_{i \in I} \psi_i$ ~
$\iff \foralli~ \pi_n(s) \models \psi_i$ ~
$\iff \foralli~ s \models cut_n(\psi_i)$ (structural induction)~
$\iff s \models cut_n(\bigwedge_{i \in I} \psi_i)$ (definition of $cut$)

\item $\varphi = \neg \psi$: \\
$\pi_n(s) \models \neg \psi$ ~
$ \iff \pi_n(s) \not \models \psi$ ~
$ \iff s \not \models cut_n(\psi)$ (structural induction)~
$ \iff s \models \neg cut_n(\psi)$ $\iff s \models cut_n(\neg \psi)$ (definition of $cut$)
\end{itemize}
\end{proof}

\begin{theorem}
Suppose $\hmeq$ is a process equivalence induced by some $\hmo \subseteq \hmli$ and compositional w.r.t.\ all projection operators $\pi_n$. AIP is sound for $\hmeq$ if and only if $\hmeq$ can be defined with some $\hmo_1 \subseteq \hmlbd$.
\end{theorem}

\begin{proof}
"$\Leftarrow$": That definability of an equivalence with a sublogic of $\hmlbd$ implies soundness of $AIP$ has been already proven in Thm.~\ref{thm:aip}.
\\ \\
"$\implies$": We have to prove that soundness of AIP implies $\exists \hmo_1 \subseteq \hmlbd:~ s \hmeq q \iff s \hmeqone q$. The desired $\hmo_1$ is constructed by applying the $cut_n$ functions  to formulas from $\hmo$: $\hmo_1 = \bigcup_{n \in \nat} \{ cut_n(\varphi) ~\mid~ \varphi \in \hmo \}$. We have:\\
$s \hmeq q~$~ $\iff \forall {n \in \nat}(\pi_n(s) \hmeq \pi_n(q))$ (soundness of AIP for $\hmeq$ \& compositionality w.r.t.\ projection)\\
$\iff \forall {n \in \nat} (\forall \varphi \in \hmo~ \pi_n(s) \models \varphi \iff \pi_n(q) \models \varphi)$\\
$\iff \forall {n \in \nat} (\forall \varphi \in \hmo~ s \models cut_n(\varphi) \iff q \models cut_n(\varphi))$ (Lem.~\ref{lem:cut})\\
$\iff \forall {n \in \nat} (\forall \psi \in \hmo_1~ s \models \psi \iff q \models \psi)$ (def. of $\hmo_1$)\\
$\iff \forall {n \in \nat} ~(\pi_n(s) \hmeqone \pi_n(q))$~
$\iff s \hmeqone q$
\end{proof}

\bibliographystyle{eptcs}

\end{document}